\documentclass{ieeetj}

\usepackage{graphicx}
\usepackage{subcaption}
\usepackage{graphicx}
\usepackage{url}
\usepackage{amsmath}
\usepackage{amssymb}
\usepackage{acronym}
\usepackage{cite}
\usepackage{comment}
\usepackage{balance}
\usepackage{acronym}
\usepackage{multirow}
\usepackage{booktabs}
\usepackage{romannum}
\usepackage{tikz}
\usepackage{amsthm}
\usepackage{amsmath}

\usepackage{hyperref}
\hypersetup{
    colorlinks=true,
    linkcolor=blue,
    filecolor=magenta,      
    urlcolor=cyan,
    }
\usepackage{algorithm,algorithmic}
\AtBeginDocument{\definecolor{tmlcncolor}{cmyk}{0.93,0.59,0.15,0.02}\definecolor{NavyBlue}{RGB}{0,86,125}}

\def\authorrefmark#1{\ensuremath{^{\textbf{#1}}}}
\newcommand{\norm}[1]{\left\lVert#1\right\rVert}

\acrodef{BCE}{binary cross entropy}
\acrodef{Sep-TFAnet}{Separation TF Attention
Network}
\acrodef{LN}{layer normalization}
\acrodef{TF}{time-frequency}
\acrodef{DOA}{direction-of-arrival}
\acrodef{MUSIC}{multiple signal classification}
\acrodef{SRP-PHAT}{steered response power with phase transform}
\acrodef{HSDA}{hierarchical search with directivity model and automatic calibration}
\acrodef{GCC}{generalized cross correlation}
\acrodef{WDO}{W-disjoint orthogonality}
\acrodef{FCN}{fully convolutional network}
\acrodef{RIR}{room impulse response}
\acrodef{w.r.t.}{with respect to}
\acrodef{RTF}{relative transfer function}
\acrodef{iRTF}{instantaneous relative transfer function}
\acrodef{TF-DOAnet}{time-frequency direction-of-arrival net}
\acrodef{VAD}{voice activity detector}
\acrodef{MAE}{mean absolute error}
\acrodef{CMS-DOA}{CNN multi-speaker DOA}

\acrodef{DNN}{deep neural network}
\acrodef{CNN}{convolutional neural network}
\acrodef{p.d.f.}{probability density function}
\acrodef{r.v.}{random variable}
\acrodef{WSJ1}{Wall Street Journal 1}
\acrodef{SIR}{signal-to interfering ratio}
\acrodef{ULA}{uniform linear array}
\acrodef{ASR}{automatic speech recognition}
\acrodef{ROS}{robot operating system}
\acrodef{ODAS}{open embedded audition system}
\acrodef{BIU}{Bar-Ilan University}
\acrodef{HWU}{Heriot-Watt University}
\acrodef{MOTA}{multiple object tracking accuracy}

\acrodef{BSS}{blind source separation}
\acrodef{DOA}{direction of arrival}
\acrodef{DC}{deep clustering}
\acrodef{DPRNN}{dual-path recurrent neural network}
\acrodef{TF}{time-frequency}
\acrodef{TCN}{temporal convolutional network}
\acrodef{ATF}{acoustic transfer function}
\acrodef{SI-SDR}{scale-invariant signal-to-distortion ratio}
\acrodef{SI-SDRi}{scale-invariant signal-to-distortion ratio improvment}
\acrodef{MSE}{mean square error}
\acrodef{DFT}{discrete Fourier transform }
\acrodef{SIR}{signal-to-interference ratio}
\acrodef{OVA}{overlap-and-add}
\acrodef{SDR}{signal-to-distortion ratio}
\acrodef{BLSTM}{bidirectional long short-term Memory}
\acrodef{SOTA}{state-of-the-art}
\acrodef{RI}{real-imaginary}
\acrodef{RIR}{room impulse response}
\acrodef{SNR}{signal-to-noise ratio}
\acrodef{RNN}{recurrent neural networks}

\acrodef{DNN}{deep neural network}
\acrodef{DOA}{direction of arrival}
\acrodef{RTF}{relative transfer function}
\acrodef{SDR}{signal direction recognition}
\acrodef{SIR}{signal to interference}
\acrodef{ATF}{acoustic transfer function}
\acrodef{STFT}{short-time Fourier transform}
\acrodef{iSTFT}{inverse short-time Fourier transform}
\acrodef{RT}{reverberation time}
\acrodef{GEVD}{generalized eigenvalue decomposition}
\acrodef{LCMV-BF}{linearly constrained minimum variance beamformer}
\acrodef{BF}{beamformer}
\acrodef{LCMV}{linearly constrained minimum variance}
\acrodef{SNR}{signal-to-noise ratio}
\acrodef{PSD}{power spectral density}
\acrodef{EVD}{eigenvalue decomposition}
\acrodef{SVD}{singular value decomposition}
\acrodef{VAD}{voice activity detector}
\acrodef{MCCSD}{multi-channel current speakers activity detector}
\acrodef{CW}{covariance-whitening}
\acrodef{FC}{fully-connected}
\acrodef{ADAM}{adaptive moment estimation}
\acrodef{ReLU}{rectified linear unit}
\acrodef{CCE}{categorical cross-entropy}
\acrodef{WFS}{waiting frame set}
\acrodef{ILRMA}{independent low-rank matrix Analysis}
\acrodef{MVDR}{minimum variance distortionless response}
\acrodef{SRP}{steered response power}
\acrodef{MCCD}{multi-channel concurrent
speakers detector}
\acrodef{NN}{neural network}
\acrodef{CNN}{convolutional neural network}
\acrodef{TF}{time frequency}
\acrodef{CSD}{concurrent speaker detector}
\acrodef{STOI}{short-term objective intelligibility}
\acrodef{PESQ}{perceptual evaluation of speech quality}
\acrodef{uPIT}{utterance-level permutation invariant training}
\acrodef{SWA}{stochastic weight averaging}
\acrodef{SGD}{stochastic gradient descent}
\acrodef{DRR}{direct-to-reverberation ratio}
\acrodef{WER}{word error rate}
\acrodef{CASA}{computational auditory scene analysis}
\acrodef{WPE}{weighted prediction error}
\acrodef{SuDoRmRf}{successive downsampling and resampling of multi-resolution features}
\acrodef{PReLU}{pre-exponential linear unit}
\acrodef{PIT}{permutation invariant training}
\acrodef{DC}{deep clustering}
\acrodef{TF}{time-frequency}

\usepackage{orcidlink}

\begin{document}

\markboth{Single-Microphone Speaker Separation and Voice Activity Detection in Noisy and Reverberant Environments}{Opochinsky {et al.}}

\title{Single-Microphone Speaker Separation and Voice Activity Detection in Noisy and Reverberant Environments}

\author{Renana Opochinsky\authorrefmark{1}\orcidlink{0009-0002-6190-7469}, Mordehay Moradi\authorrefmark{1}\orcidlink{0009-0000-1190-0555}, and Sharon Gannot\authorrefmark{1}\orcidlink{0000-0002-2885-170X}, Fellow IEEE}
\affil{Bar-Ilan University,
Ramat-Gan, 5290002
ISRAEL}
\corresp{Corresponding author: Sharon Gannot (email: sharon.gannot@ biu.ac.il).}
\authornote{The project has received funding from the European Union’s Horizon 2020 Research and Innovation Programme, Grant Agreement
No. 871245; and was also supported by the Israeli Ministry of Science \& Technology.\\
Renana Opochinsky and Mordehay Moradi have equally contributed to the paper.
}

\begin{abstract}
Speech separation involves extracting an individual speaker's voice from a multi-speaker audio signal. The increasing complexity of real-world environments, where multiple speakers might converse simultaneously, underscores the importance of effective speech separation techniques.
This work presents a single-microphone speaker separation network with TF attention aiming at noisy and reverberant
environments. We dub this new architecture as \ac{Sep-TFAnet}. In addition, we present a variant of the separation network, dubbed \ac{Sep-TFAnet}\textsuperscript{VAD}, which incorporates a \ac{VAD} into the separation network.

The separation module is based
on a \ac{TCN} backbone
inspired by the Conv-Tasnet architecture with multiple modifications. Rather than a
learned encoder and decoder, we use  \ac{STFT} and \ac{iSTFT} for the analysis and synthesis, respectively. Our system is specially developed for human-robotic interactions and should support online mode. 
The separation capabilities of Sep-TFAnet\textsuperscript{VAD} and \ac{Sep-TFAnet} were evaluated and extensively analyzed under several acoustic conditions, demonstrating their advantages over competing methods. 
Since separation networks trained on simulated data tend to perform poorly on real recordings, we also demonstrate the ability of the proposed scheme to better generalize to realistic examples recorded in our acoustic lab by a humanoid robot.
Project page: \url{https://Sep-TFAnet.github.io} 

\end{abstract}



\maketitle

\section{Introduction}


\IEEEPARstart{B}{lind} speaker separation is a rapidly growing research field aiming to separate a mixture of speech signals into their components. This technology has important applications in robot audition, speech recognition, hearing aids, and telecommunications. The introduction of \ac{DC}~\cite{hershey2016deep} and \ac{PIT}~\cite{yu2017permutation} has sparked increasing interest in single-microphone speaker separation algorithms that rely on \acp{DNN}.

Many recent separation schemes are directly applied in the time domain using learnable encoder-decoder structure, e.g., Conv-Tasnet \cite{luo2019conv}, \ac{DPRNN} \cite{luo2020dual} and many more \cite{zeghidour2021wavesplit,zhao2023mossformer,nachmani2020voice,lutati2022sepit}. 

The Conv-Tasnet algorithm, as introduced in \cite{luo2019conv}, employs one-dimensional convolutional layers to infer self-learned representations of the input signal. These embedded-domain representations are utilized to compute masks for each individual speaker, which are subsequently applied to separate the speakers. This approach employs the \ac{SI-SDR} \cite{le2019sdr} loss function, specifically designed to capitalize on temporal information. The algorithm achieves improved performance compared with \ac{SOTA} methods, probably because its features are learned rather than manually designed.

 A transformer-based neural network architecture, dubbed SepFormer, was presented \cite{subakan2021attention} and \cite{subakan2022using}. The \ac{SuDoRmRf}, presented in \cite{tzinis2022compute}, is based on simple one-dimensional convolution layers requiring significantly less memory and parameters than the existing aforementioned methods.
Several challenges and limitations remain unsolved in current methods. Many of them perform poorly in the presence of background noise and high reverberation, resulting in significant degradation in the quality of the separated signals. Robustness to noise and reverberation is a key challenge in real-world applications of speech source separation.
These approaches mainly focused on a non-realistic anechoic dataset as WSJ-2mix \cite{hershey2016deep} and WHAM! \cite{wichern2019wham}. Specifically, the frame length is relatively short in time domain approaches, allowing proper separation only in an anechoic environment \cite{cord2022monaural}, \cite{heitkaemper2020demystifying}.



In the current contribution, we propose the \ac{Sep-TFAnet}, a modified separation approach that leverages the Conv-Tasnet \cite{luo2019conv} backbone, using \ac{TCN}, toward a more realistic scenario. By incorporating an attention layer and training with reverberant references, the proposed network improves the separation scores in more challenging environments compared to other competing methods. 
In addition, we propose \ac{Sep-TFAnet}\textsuperscript{VAD}, which simultaneously infers all speakers' activity patterns. In other words, together with the separation, we jointly train a \ac{VAD} network. \ac{VAD} is an essential component in multiple downstream tasks, e.g., post-filters, beamforming, and localization. Note that several papers combine VAD with multi-channel speech processing \cite{wang2021continuous}, \cite{lin2021sparsely}, and for extraction \cite{lin2021sparsely}.

One of the main objectives of \ac{Sep-TFAnet}\textsuperscript{VAD} is to improve a robot's auditory capabilities. 
Developing socially assistive robots that can perform multi-person interactions requires progress in speech-related tasks such as speech source separation, localization, etc. The robot is supposed to function in a crowded environment. Therefore, we trained and adapted our system to complex acoustic conditions. 
In this context, the \ac{VAD} can be utilized as a voice activity controller, assisting the robot in engaging in human-like interactions.
The \ac{VAD} decisions can facilitate applying a post-filtering operation to suppress further residual interfering signals and noise and enable the estimation of several building blocks of subsequent multichannel processing, e.g., an \ac{LCMV} beamformer.


In the proposed network, inspired by Conv-Tasnet \cite{luo2019conv}, we substitute the learnable time domain encoder-decoder with an \ac{STFT} to better deal with high reverberation scenarios. In \cite{cord2022monaural}, it was demonstrated that the \ac{STFT} is a more valid choice for the encoder and decoder stages than the learnable representations. It is also demonstrated that jointly addressing
both reverberation and overlapped speech appears to be a challenging task. 

Furthermore, this paper introduces a new simulated dataset with more realistic conditions, such as the typical overlap between speakers and higher perceived reverberation level than the WHAMR! dataset \cite{maciejewski2020whamr}. Moreover, we also introduce real-world data recorded on a humanoid robot. Our model performs
comparably or better than \ac{SOTA} models on this new, more challenging dataset.

The main contributions of this paper are:
1)  A more realistic dataset including ``real-world" experiments; 2) An elaborate network based on Tasnet that outperforms \ac{SOTA} methods in challenging scenarios; 3) An `online mode' for low latency applications in practical setups; 4) An extensive analysis of the separation results; and 5) A \ac{VAD} that operates in parallel to the separation module and achieves accurate activity readings.

The paper is organized as follows. In Sec.~\ref{sec:prob}, we formulate
the separation problem.   
In Sec.~\ref{sec:prop}, we describe the proposed model, \ac{Sep-TFAnet}, and its components. In Sec.~\ref{sec:exp},
we demonstrate the performance of the proposed algorithm through an extensive simulation study and real-world recordings. A comparison
between the \ac{Sep-TFAnet} (with and without VAD) and the competing algorithms is carried out. Section~\ref{sec:conc} concludes the paper.

\begin{figure}[htbp]
\centering
\includegraphics[width=\columnwidth]{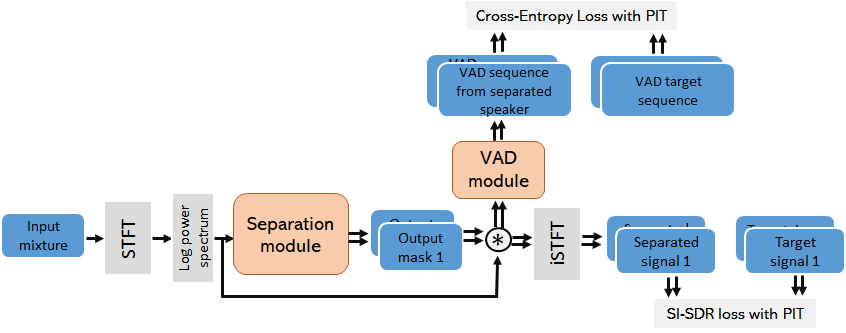}
 \caption{\ac{Sep-TFAnet}\textsuperscript{VAD} architecture. Learnable blocks are depicted in orange, and data blocks are in blue.} 
\label{fig:arch}
\end{figure}
\section{Problem Formulation}
\label{sec:prob}

Let $x(t)$ be a mixture of $I$ concurrent speakers captured by a single microphone:
\begin{equation}
    x(t)=\sum^I_{i=1}  \{s_i\ast {h}_i\}(t) +n(t) \quad {t=0,1,\ldots,T-1},
    \label{eq:mix_time}
\end{equation}
where $s_i(t)$ represents  the signal of the $i$-th speaker, ${h}_i(t)$ represents the, possibly time-varying, 
\ac{RIR} between the $i$-th speaker and the microphone, and $n(t)$ represents an additive noise. Time variations of the \ac{RIR} can be attributed to either the sources' movements, the microphone, or both. In this paper, we deal with static scenarios.
In the \ac{STFT} domain, and under the assumption of sufficiently long time frames, \eqref{eq:mix_time} can be reformulated as, 
\begin{equation}
x(l,k) =  \sum^I_{i=1} s_i(l,k)  h_i(l,k) +n(l,k),
\label{eq:mix_STFT}
\end{equation}
where $l \in \{0,\ldots, L-1\}$ and $k \in \{0,\ldots, K-1\}$ are the time-frame and the frequency-bin indexes, respectively, with $L$ the total number of time-frames and  $K$ the total number of frequency bins. In our study, we only address the case $I=2$. 
Denote the outputs of the separator system in the \ac{STFT} domain as $\hat{s}_1(l,k)$ and $\hat{s}_2(l,k)$.

We assume a fully-supervised setting, in which the goal is to infer a model that separates two output signals $\hat{s}_1(t)$ and $\hat{s}_2(t)$ from an unseen mixture $x(t)$, by maximizing the \ac{SI-SDR} between the separated signals and the reverberated clean signals, $\{s_1\ast {h}_1\}(t)$ and $\{s_2\ast {h}_2\}(t)$, respectively.

In addition, in one version of our system \ac{Sep-TFAnet}\textsuperscript{VAD}, the activity of each separated speaker
per time frame is determined by applying a \ac{VAD} network. This task also assumes a fully-supervised training.


\begin{figure}[htbp]
  \centering
  \begin{subcaptiongroup}
    \centering
    \parbox[b]{.45\textwidth}{%
    \centering \includegraphics[width=0.95\columnwidth]{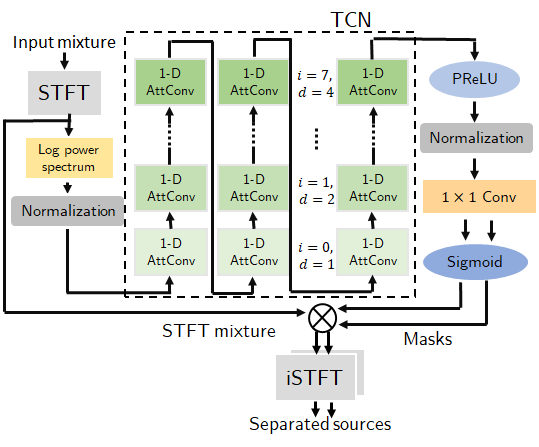}
        \caption{Separation module 
        }
\label{fig:sep_net}
        }
    \parbox[b]{.45\textwidth}{%
    \centering
    \includegraphics[width=0.42\columnwidth]{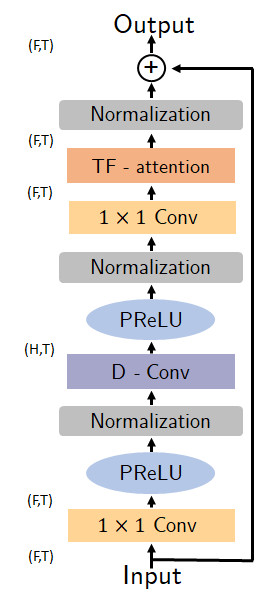} 
        \caption{1-D AttConv}
        \label{fig:1-D-Conv}}
    \parbox[b]{.45\textwidth}{%
    \centering   \includegraphics[width=0.5\textwidth]{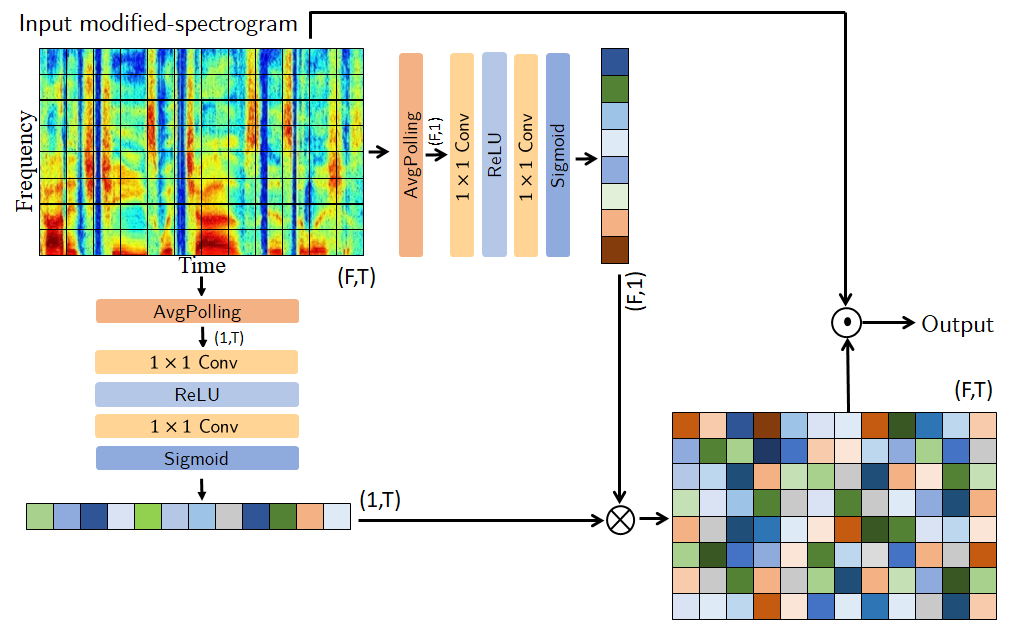}
        \caption{TF attention layer}
        \label{fig:TF_att}}       
  \end{subcaptiongroup}  
  \caption{Separation Module: Architecture.}
  \label{fig:net_arch}
\end{figure}

\section{Proposed Model}
\label{sec:prop}

The model comprises two main components: a separation module and a \ac{VAD} module. The separation module is based on a \acf{TCN} backbone \cite{lea2016temporal}. Rather than a learned encoder and decoder, we use the \ac{STFT} and the \ac{iSTFT}. As demonstrated in \cite{cord2022monaural,wang2022stft}, audio processing algorithms that are based on \ac{STFT}-\ac{iSTFT} are advantageous in high reverberation as compared with learned encoder-decoder.  

The \ac{VAD} module determines the activity patterns of each separated signal and can be useful in downstream tasks.
A block diagram of the entire system is depicted in Fig.~\ref{fig:arch}.

Next, we elaborate on the various components of the system. 

\subsection{Separation Module}
%
The Separation Module is the main component of our proposed scheme. The various components of this module are depicted in Fig.~\ref{fig:net_arch}.

An \ac{STFT} first analyzes the raw audio signal. Then, the log-spectrum is calculated, followed by \ac{LN}. The result of these operations constitutes the input to the separation network.
%
The main processing module is an adapted \ac{TCN} (dashed part in Fig.~\ref{fig:sep_net}). Originally, the \ac{TCN} is a series of identical 1-D Conv blocks with increasing dilation factors and with zero-padding along the time dimension to maintain their integrity. The dilation factor enables capturing a sufficiently long temporal context of the speech signal. We stress that past \ac{STFT} frames are also relevant for the separation task in high reverberation levels and should be considered. Here, the adapted \ac{TCN}  consists of three repeats of a stack of eight 1-D AttConv blocks, as proposed in \cite{luo2019conv} and depicted in Figs.~\ref{fig:sep_net}, \ref{fig:1-D-Conv}. 

Unlike the original 1-D Conv blocks in the Conv-Tasnet algorithm \cite{luo2019conv}, our 1-D AttConv blocks have only one output and do not feature an additional skip connection output from the output of each block to the overall \ac{TCN} output (see Fig.~\ref{fig:1-D-Conv}). We have decided to avoid this additional output since according to empirical evidence indicated in \cite{ravenscroft2023deformable} and affirmed by our findings, they do not improve performance but rather increase the number of parameters. 

For each 1-D AttConv block (see Fig.~\ref{fig:1-D-Conv}), the input of the block is summed to its output. Additionally, since the input to our network is an \ac{STFT} representation, the dilation factors were chosen to maintain as small as possible receptive field but long enough to address reverberation. As a suitable compromise, we set $d = (i \bmod 4) + 1,\,  0 \le i \le 7 $, where $d$ is the dilation factor and $i$ denotes the number of the 1-D AttConv block in each repeat as depicted in Fig.~\ref{fig:sep_net}.
The first $1 \times 1$  Conv layer in each 1-D AttConv block has a kernel size 1. The number of filters is set to $F$, the number of frequency bins, to capture the frame-wise frequency patterns. In addition, following the $1 \times 1 $ Conv layer, the D-Conv layer expounded upon in \cite{luo2019conv}, is subsequently applied with $H$ output filters, $H>F$ with the purpose of achieving a richer representation while concurrently minimizing the number of parameters. 

Another component contributing to the network's performance is the \ac{TF} attention block adopted from \cite{9966661}, which was proposed in the context of noise reduction. The TF attention block is depicted in Fig.~\ref{fig:TF_att}. This block is positioned after the final $1 \times 1 $ Conv layer of each 1-D AttConv block, followed by a normalization layer, to optimize further the network's ability to learn to recognize complex patterns in speech data.
Despite its relatively small number of parameters, this module improves the performance by approximately 0.5~dB. As depicted in Fig.~\ref{fig:TF_att}, the block commences with average pooling layers on both the time and frequency dimensions, followed by $1 \times 1 $ Conv layers and activation functions. These layers generate an attention mask, subsequently multiplied element-wise with the modified spectrogram of the input signal.


Finally, after applying all 1-D AttConv blocks, a \ac{PReLU} activation function, along with a \ac{LN} and a $1 \times 1 $ Conv, is employed to estimate the masks, which are subsequently passed through a Sigmoid activation function, which output is confined to the interval $[0,1]$ as depicted in Fig.~\ref{fig:net_arch}.
In conjunction with the \ac{STFT}, this activation function preserves the audio scale of the output along the utterance and can be beneficial in online mode.
The estimated speakers' signals are transformed back to the time domain by augmenting the masked spectrum with the noisy and mixed input signal phase and then applying the \ac{iSTFT}.

\vspace{-0.5\baselineskip}
\subsection{Online Mode}
Developing an online variant of the separation algorithm is vital in human-robot communication, as it may facilitate proper interaction between the robot and the human speaker. We employ a distinctive strategy instead of processing the complete utterance in batch mode.
We divide the input into short segments comprising past, present, and future samples. We apply the separation operation to each segment by employing a sliding window. To maintain consistency of the separated signals between all segments, we apply \ac{PIT} with $L1$ loss.
Notably, the step size for the aforementioned sliding window technique is set to be equal to the number of future samples.
\vspace{-0.5\baselineskip}
\subsection{\Acf{VAD} network}

The purpose of the \ac{VAD} network is to infer the activity patterns of the separated speakers.
The inputs to the network are the extracted masks from the jointly trained separation module.
The \ac{VAD} network consists of a 1-D convolution layer with four filters, followed by a \ac{PReLU} activation function and a normalization layer. The activity patterns of the corresponding speakers are finally obtained by applying a 1-D convolutional layer with one filter, followed by a hard threshold to obtain a binary activity decision for each frame. In addition, we modify the 1-D AttConv block, as proposed in \cite{liu2020rethinking}, such that the output of 1-D AttConv block is:
\begin{equation}
    \textrm{output}\textsuperscript{1-D AttConv} = \textrm{LN}(x + \textrm{LN}(x + y_{\textrm{att}}))
\end{equation}
where $x$ is the input to the AttConv block, and $y_{\textrm{att}}$ is the output of the TF-attention layer within the AttConv block.
\vspace{-0.5\baselineskip}
\subsection{Objective Functions}

We experimented with several separation objective functions to efficiently train the model and found that the \ac{SI-SDR} loss yields the best perceptual improvement. 
The loss is formulated as follows: 
\begin{equation}
\textrm{SI-SDR}\left( s,\hat{s} \right)= 10 \log_{10} \left( \frac{\norm{\frac{\langle {\hat{s},s} \rangle}{\langle {s,s} \rangle} s}^2}{\norm{\frac{\langle {\hat{s},s} \rangle}{\langle {s,s} \rangle} s-\hat{s}}^2} \right).
\end{equation}
To alleviate the permutation problem, common to separation problems, \ac{uPIT} \cite{kolbaek2017multitalker} was employed.
We stress that the target signals during training were the reverberant signals, namely the anechoic signals convolved with the corresponding \acp{RIR}. Hence, the signal focuses on the separation task and does not attempt to dereverberate the separated signals. While this may improve separation scores, the performance of \ac{ASR} systems may deteriorate in high reverberation levels. If reverberation is an issue, a dereverberation module, e.g., \cite{yemini2020scene}  or \cite{delcroix2014linear}, can be applied as a post-filter. 

For training the \ac{VAD} module, we have used the \ac{BCE} loss:
 \begin{equation}
 \textrm{BCE}=
 -\sum_{i=1}^I \sum_{l=0}^{L-1} v_i(l)\log(p_{i}(l)) 
 + (1 -  v_i(l) )\log(1 - p_{i}(l)),
\end{equation}
 where $p_{i}(l)$ is the output of the VAD network indicating the probability of speaker $i=1,\ldots,I$ at time frame $l$ to be active, ${v}_{i}(l)$ is the true activity of speaker $i$ at time frame $l$.


\section{Experimental Study}
\label{sec:exp}
\subsection{Datasets}
The WHAMR! dataset \cite{maciejewski2020whamr} is widely used for speech separation in reverberant environments. According to our tests, the reported reverberation level, $ 0.1-1$~s, is inaccurate, and in practice, it is perceived as less reverberant than reported.\footnote{Analyzing the WHAMR! recipe for Pyroomacoustics, we believe that this discrepancy results in from the definition of the reflection coefficient.} We have, therefore, constructed our own database with another simulator and used it for training and evaluation in conjunction with the WHAMR! dataset.  

To generate our dataset, clean speech utterances were drawn from the clean corpus of the Librispeech database \cite{panayotov2015librispeech} and then convolved with \acp{RIR} generated using the image method \cite{allen1979image} implemented in\cite{habets2006room}. The reverberation level is uniformly set in the $[0.2, 0.6]$~s range. The reverberated signals were then mixed with SIR=0~dB and contaminated with babble noise from the WHAM! dataset \cite{wichern2019wham} with SNR uniformly drawn in the range $[0, 15]$~dB. The speakers were located in random positions (under physical constraints) inside shoebox rooms with random dimensions. The room's length and width were uniformly drawn in the range $[4.5,6.5]$~m, and the room's height was uniformly drawn in the range $[2.5,3]$~m. The speakers are partially overlapping, with overlap percentage randomly selected from $[50\%, 75\%, 100\%]$. The length of each sample is 10~Sec.
Overall, our simulated data consists of 155 hours of audio samples for train, 39 hours for validation, and 22 hours for testing. 
 
\begin{figure}[htbp]
  \centering
  \begin{subcaptiongroup}
    \centering
    \parbox[b]{.45\textwidth}{%
    \centering \includegraphics[width=0.35\textwidth]{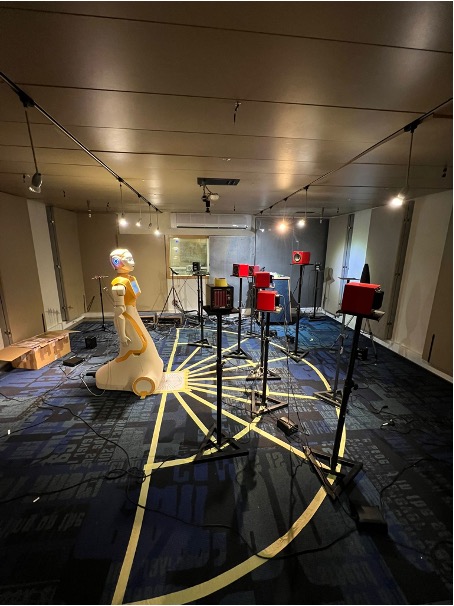}
    \caption{Recording setup with ARI at BIU acoustic lab. ARI was positioned at the center of the acoustic lab, with a set of loudspeakers in front of it, on two semi-circles with approximately 1~m and 2~m radius, respectively. In our experiments, we only used the inner semi-circle with five loudspeakers positioned at $[-65, -30, 0, 30, 65]^\circ$. The speech signals were simultaneously played from two randomly chosen loudspeakers. The lab's computer automatically controlled the entire scenario.
    }
    \label{fig:lab}}

    \parbox[b]{.45\textwidth}{%
    \centering
    \includegraphics[width=0.45\textwidth]{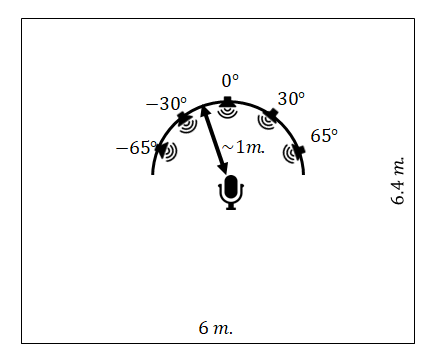}
    \caption{Geometric setup of the experiment at BIU Acoustic Lab.}
    \label{fig:lab_scheme}}
        
  \end{subcaptiongroup}
  
 \caption{Experimental setup with ARI robot in the center of the Bar-Ilan acoustic lab with loudspeakers in a semi-circle at the front of the robot.}
    \label{fig:exp_setting}
\end{figure}

\vspace{-0.5\baselineskip}
\subsection{Robot Interaction: Experimental Setup}
To further analyze the algorithm's performance in a real-world setting, specifically in the context of robot audition, a recording campaign was held in the acoustic laboratory at Bar-Ilan University. This lab is a $6 \times 6 \times 2.4$~m room with a reverberation time controlled by 60 interchangeable panels covering the
room facets. 
In our experiments, the reverberation time was set (by changing the panel arrangements) to either 350 ms, typical of a meeting room, or 600 ms, typical of a lecture hall. The room layout is depicted in Fig.~\ref{fig:lab}.

ARI\footnote{\url{https://pal-robotics.com/robots/ari/}} by PAL Robotics is a humanoid robot with an Intel i9 processor and Nvidia Orin GPU. The goal of our project is to improve the social capabilities of the robot and specifically to develop its capabilities to engage with patients in a hospital environment.\footnote{SPRING - Socially Pertinent Robots in Gerontological Healthcare, a project under the Horizon2020 programme, \url{https://spring-h2020.eu/}.} 

ARI is equipped with ReSpeaker 4-Microphone Array v2.0, \footnote{\url{https://wiki.seeedstudio.com/ReSpeaker_Mic_Array_v2.0/}} installed inside the robot's compartment, 80~cm above its base. 
We only used one of the microphones for evaluating the single-microphone algorithms.

In our experimental setup, ARI was positioned at the center of the acoustic lab, with a set of loudspeakers in front of it, on two semi-circles with approximately 1~m and 2~m radius, respectively. Our experiments only used the inner semi-circle with five loudspeakers positioned at $[-65, -30, 0, 30, 65]^\circ$. To generate a sample, we randomly selected two loudspeakers and played speech utterances randomly drawn from 
the Librispeech test set \cite{panayotov2015librispeech}. The utterances were separately recorded by ARI and then manually mixed to enable \ac{SI-SDR} calculation. The overlap between the speakers was randomly set in the range $[25\%, 50\%]$. No external noise was added to the recordings; hence, only sensor and low-level ambient noise are present. Overall, 200 samples were generated at each reverberation level. The experimental setup is schematically depicted in Fig.~\ref{fig:lab_scheme}.


\vspace{-0.5\baselineskip}
\subsection{Training Procedure}
In this section, we give the technical details of the training procedure. The model was trained on GPU servers, Tesla V100 SXM2 32GB,  using 315
 epochs. A $1e^{-3}$ learning rate was chosen, coupled with the Adam optimizer. During the training process, gradient clipping is applied with a maximum L2-norm of 5 to ensure stability and prevent exploding gradients. 
The model has approximately 5M trainable parameters.
The parameters of the \ac{STFT} and hyperparameters of the network are detailed in Table~\ref{table:hyperparameters}.

\begin{table}[htbp]
\caption {The parameters of the \ac{STFT} and hyperparameters of \ac{Sep-TFAnet}\textsuperscript{VAD}.} \label{table:hyperparameters} 
\centering
\begin{tabular}{llc}
\toprule
           & Variable   & Value   \\
\midrule
\multirow{4}{8em}{STFT}           & Hop length       & 256    \\
                                &  FFT bins      & 512    \\
                              & Window & Hamming   \\
                              & Window length & 512  \\
                              \midrule
\multirow{4}{8em}{Hyperparameter} & H          & 512     \\
                                & F          & 256     \\
                                & $F_s$        & 16~kHz      \\
                                & Batch size &   16      \\
                                \bottomrule
\end{tabular}
\end{table}

\vspace{-0.5\baselineskip}
\subsection{Baseline Methods}
We compared the proposed method to two competing methods that have a comparable number of parameters, the SuDoRmRf \cite{tzinis2020sudo,tzinis2022compute} and the Conv-Tasnet \cite{luo2019conv}. \ac{SuDoRmRf} model leverages the effectiveness of iterative temporal resampling strategies to avoid the need for multiple stacked dilated convolutional layers. Another baseline method is the Conv-Tasnet algorithm \cite{luo2019conv}, as it shares a similar backbone with our separation module, namely the \ac{TCN} module. The Conv-Tasnet is applied in the time
domain with learned encoder-masking-decoder architecture. 
As mentioned previously, our approach diverges from the original Conv-Tasnet method in several important aspects.
The Conv-Tasnet and our networks have a similar number of parameters $\sim5.5$ million, while the \ac{SuDoRmRf} model is more compact with  $\sim2.7$ million parameters. We trained the baselines with our new simulated dataset for the results reported in the current paper.

\vspace{-0.5\baselineskip}
\subsection{Experimental Results}
\subsubsection{Separation with Simulated Data}

The \ac{SI-SDR}, averaged over all utterances, for \ac{Sep-TFAnet}, \ac{Sep-TFAnet}\textsuperscript{VAD} and competing algorithms for both the WHAMR! database and the new simulated data are depicted in Table~\ref{table:sisdr_results}. 
\begin{table}[htbp]
\centering 
\caption {Averaged \ac{SI-SDR} [dB]} \label{table:sisdr_results} 
\centering
\begin{tabular}{lcc} 
\toprule
    Algorithm  & WHAMR! & Simulated Data   \\ 
\midrule
    Unprocessed (mix)  &   -0.86 & -2.04 \\
    \ac{Sep-TFAnet}  &  8.1 & \textbf{6.92} \\
    \ac{Sep-TFAnet}\textsuperscript{VAD}  &  7.86 & 6.77   \\
     \ac{SuDoRmRf} &  7.04   & 5.6  \\  
    Conv-Tasnet &  \textbf{8.3}   & 6.6  \\  
    \bottomrule
\end{tabular}    
\end{table}
The \ac{SI-SDR} scores of the proposed method and the Conv-Tasnet are comparable, while the \ac{SuDoRmRf} performance is lower. 
It is also evident that the new simulated data is more challenging than the WHAMR! database.


\begin{figure}[htbp]

  \centering
  \begin{subcaptiongroup}
    \centering

\parbox[b]{.24\textwidth}{%
    
    \includegraphics[width=0.25\textwidth]{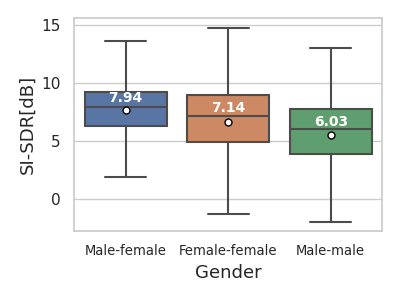}
        \caption{SI-SDR vs. Gender.}
        \label{fig:sisdr_gender}}
 \hfill
\parbox[b]{.24\textwidth}{%
    
    \includegraphics[width=0.25\textwidth]{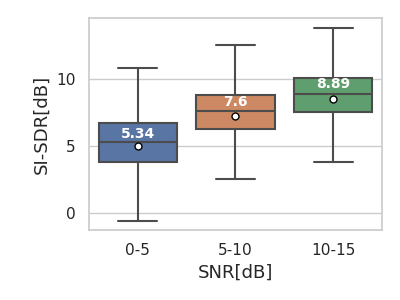} 
        \caption{SI-SDR vs.~SNR.}
        \label{fig:sisdr_snr}}

    \parbox[b]{.24\textwidth}{%
    
    \includegraphics[width=0.25\textwidth]{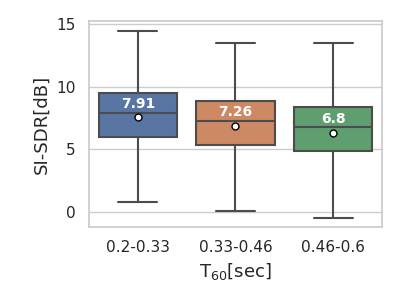}  
        \caption{SI-SDR vs.~$T_{60}$.\newline}\label{fig:sisdr_T60}}
 \hfill
    \parbox[b]{.24\textwidth}{%
    
    \includegraphics[width=0.25\textwidth]{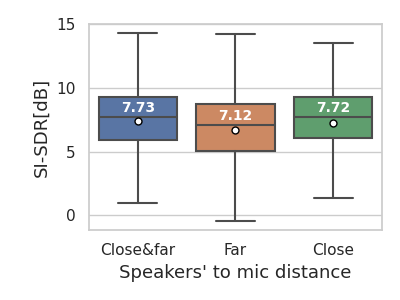} 
        \caption{SI-SDR vs. \\speakers' to mic distance} 
        \label{fig:sisdr_drr}}

  \end{subcaptiongroup}

    \caption{The \ac{SI-SDR} of the proposed algorithm tested on our new simulated dataset and categorized according to speakers' gender, \ac{SNR} level, reverberation level, and speakers-to-mic distance. The number indicates the median value, while the mean score is indicated by $\circ$.}
    \label{fig:SI-SDR}
\end{figure}
Further insights into the performance of \ac{Sep-TFAnet} can be inferred by carefully examining the statistics of the results.  Figure~\ref{fig:SI-SDR} depicts box plots of the \ac{SI-SDR} categorized according to speakers' gender, \ac{SNR} level, reverberation level, and speakers’ to-mic distance.
It is evident from Fig.~\ref{fig:sisdr_gender} that the best results are obtained for the mixed-gender case, while the worst results are for the male-male mixture.
A marginal improvement is demonstrated with increasing \ac{SNR} level, as evident from Fig.~\ref{fig:sisdr_snr}. 
The dependency on the reverberation level is indicated in Fig.~\ref{fig:sisdr_T60}, with a clear performance drop from lower levels to higher levels of $T_{60}$. 
Finally, as depicted in Fig.~\ref{fig:sisdr_drr}, best results are obtained if the distance of both sources from the microphone is lower than the critical distance (and consequently high \ac{DRR}), and worst when both sources are beyond the critical distance. It should be stressed that above the critical distance, the reverberation tail dominates the \ac{RIR}, and hence, the signal is perceived as more reverberant. 

\begin{figure}[htbp]
  \centering
  \begin{subcaptiongroup}
    \centering
    \parbox[b]{.5\textwidth}{%
    \centering \includegraphics[width=0.47\textwidth]{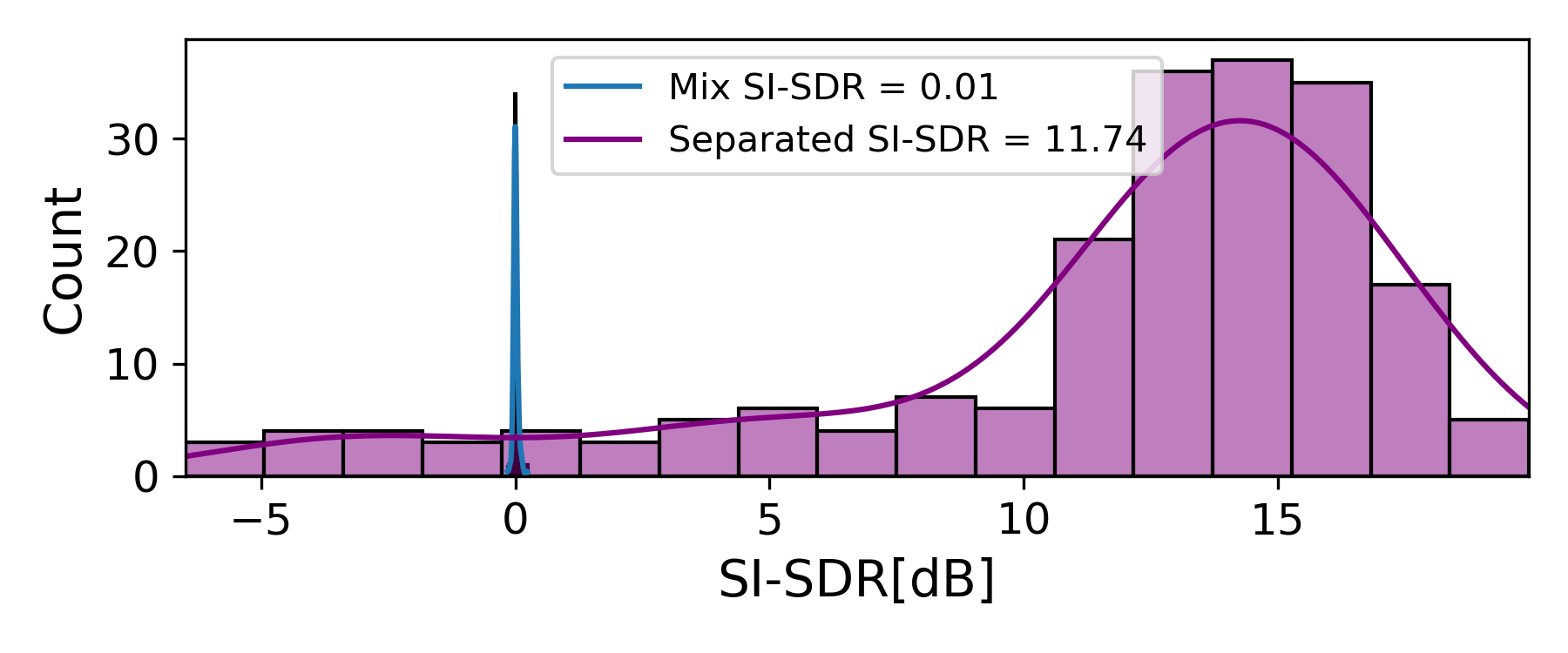}
        \caption{\ac{SI-SDR} histograms}
        \label{fig:sisdr_ari}}

    \parbox[b]{.5\textwidth}{%
    \centering
    \includegraphics[width=0.47\textwidth]{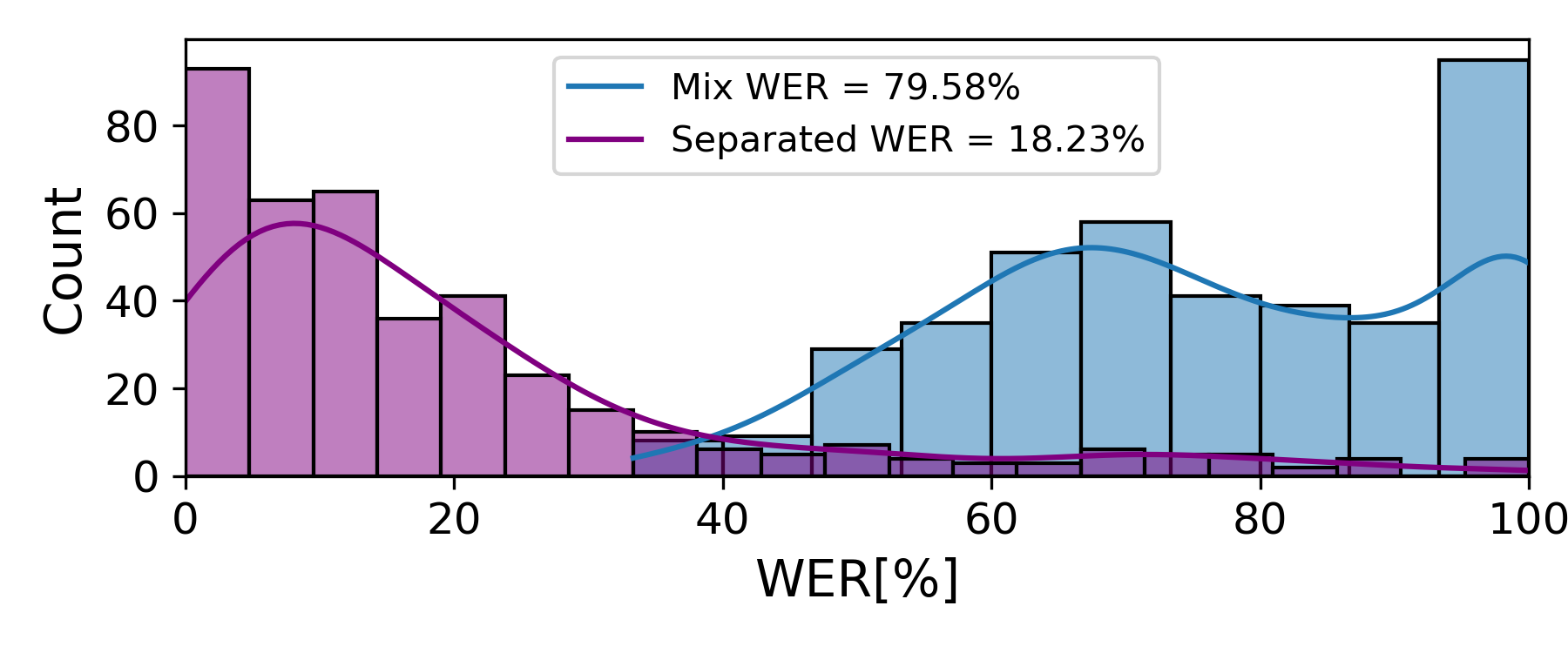} 
        \caption{\ac{WER} histograms}
        \label{fig:wer_ari}}
        
  \end{subcaptiongroup}
  
    \caption{\ac{SI-SDR} and \ac{WER} improvements with ARI recordings, $T_{60}=0.35$~s and low sensor noise. The overlap of the speakers was randomly set to the range $[25\%, 50\%]$.}
    \label{fig:ARI_res}
\end{figure}
\vspace{-3\baselineskip}

\subsubsection{Separation with Data Recorded on ARI Robot 
}

We have also used the database recorded on the robot's microphone to verify the algorithm's applicability to our scenarios.
In the network's regular mode, a significant improvement both in terms of \ac{WER} (using NVIDIA RIVA Conformer-based \ac{ASR}\footnote{\url{https://catalog.ngc.nvidia.com/orgs/nvidia/teams/tao/models/speechtotext_en_us_conformer}}) and \ac{SI-SDR} can be deduced from the histograms in Fig.~\ref{fig:ARI_res}, analyzing the results for $T_{60}=0.35$~sec and low-sensor noise scenario. The mean \ac{SI-SDR} was improved from approximately 0~dB to 11.74~dB, and the mean \ac{WER} from 76\% to 18.23\%. 

Table~\ref{table:online_offline_results} depicts the mean \ac{SI-SDR} and \ac{WER} results of this real-world experiments for the proposed algorithm (with and without \ac{VAD}), and Conv-Tasnet. We present the results for the regular mode (batch processing). The online mode will be discussed in the next section. 

\begin{figure*}[htbp]
  
   \centering
 
\includegraphics[width=\textwidth]{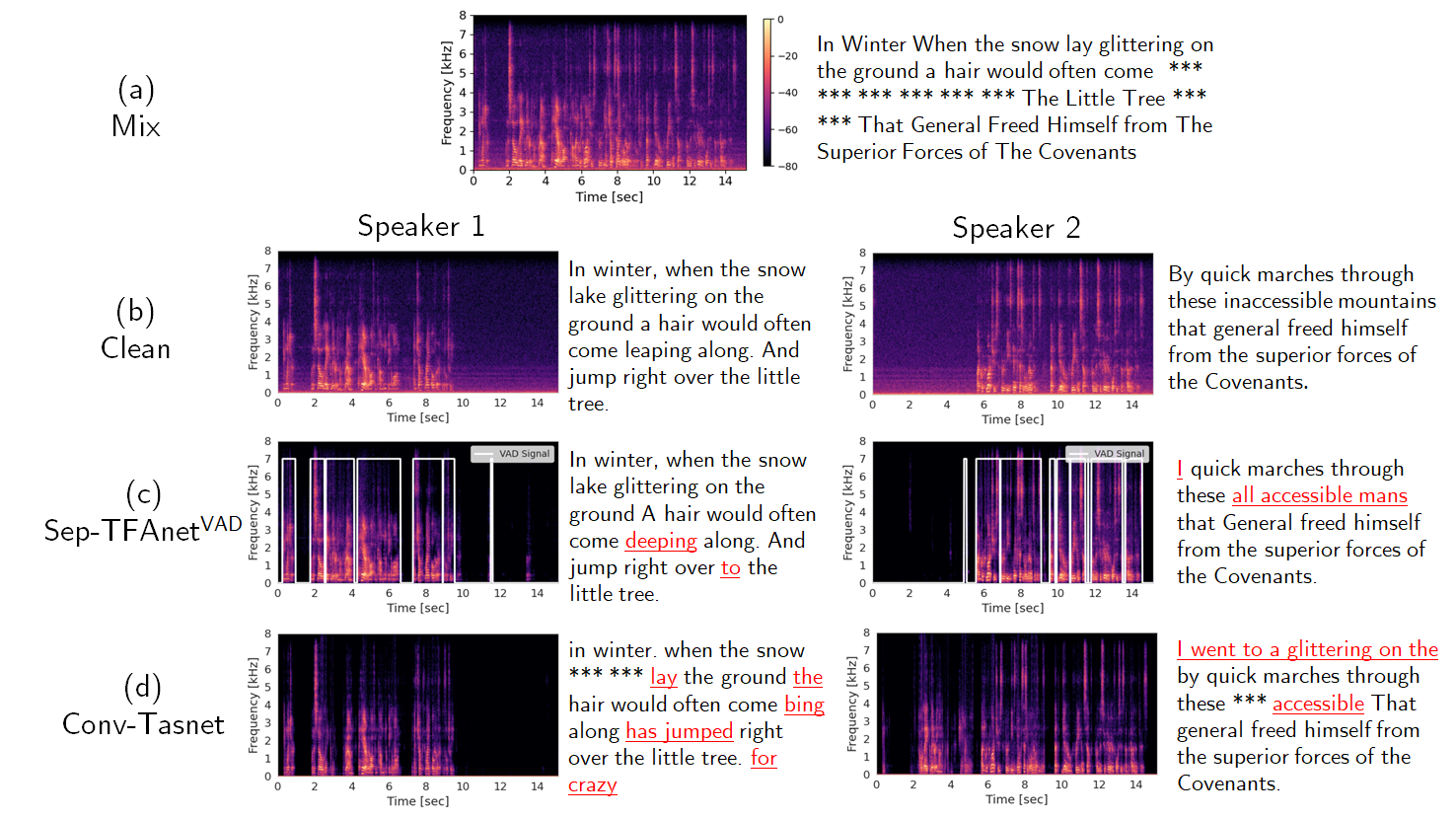}
 \caption{Spectrograms of (a) mixed-signal, (b) clean-reference signals, (c) separated outputs of our network, VAD outputs in white on our results, (d) separated outputs of Tasnet and respective transcripts. The words in red indicate wrong transcription, and `*' - indicates missing words. The reverberation level was $T_{60}=0.35$~sec with 50\% overlap between the speakers. The loudspeakers were located at $35^\circ$ and $60^\circ$.}
    \label{fig:spec_son}
\end{figure*}

In the $T_{60}=0.35$~sec case, it is evident that the proposed algorithm significantly outperforms the Conv-Tasnet and achieves much lower WER.
The performance advantages are less pronounced for the $T_{60}=0.6$~sec case, with our proposed algorithm only slightly outperforming Conv-Tasnet. We also note that \ac{Sep-TFAnet} performs better than \ac{Sep-TFAnet}\textsuperscript{VAD} in both $T_{60}$ conditions. 
\begin{table}[htbp]
\centering 
\caption {Mean separation results on ARI recorded data. 
} \label{table:online_offline_results} 
\centering
\resizebox{\columnwidth}{!}
{
\begin{tabular}{llcccc} 
\toprule
 & & \multicolumn{2}{c}{SI-SDR[dB]} &
      \multicolumn{2}{c}{WER} \\
\cmidrule(lr){3-6}
 & &\multicolumn{4}{c}{$T_{60}$ [sec]}\\
 \cmidrule(lr){3-6}
    Mode & Algorithm  & 0.35 & 0.6 & 0.35 & 0.6    \\ 
\cmidrule(lr){1-1}\cmidrule(lr){2-2}\cmidrule(lr){3-4}\cmidrule(lr){5-6}
     \multirow{3}{*}{Regular} & \ac{Sep-TFAnet}  &    \textbf{11.74} & \textbf{9} & \textbf{18.23} & \textbf{24.96} \\
      & \ac{Sep-TFAnet}\textsuperscript{VAD} &  11.55 & 8.7 & 18.69 &  26.41 \\  
    & Conv-Tasnet &  10.88 & 8.33 & 27.6 &  27.71\\  

\midrule
     \multirow{3}{*}{Online} & \ac{Sep-TFAnet}&    \textbf{11.64} & 8.4 & \textbf{18.3} & 26.51 \\
      & \ac{Sep-TFAnet}\textsuperscript{VAD} &  9.87 & 7.81 & 23.88 &  29.77 \\  
    & Conv-Tasnet &  9.88 & \textbf{8.31} & 21.32 &  \textbf{26.16}\\  

    
    \bottomrule
\end{tabular} 
}

\label{table:ASR}
\end{table}
For qualitative assessment, we give examples of separated signals with sonograms and transcriptions in Fig.~\ref{fig:spec_son}. The reverberation level was chosen as $T_{60}=0.35$~sec with 50\% overlap between the speakers' activities. The loudspeakers were located at $35^\circ$ and $60^\circ$. In this example, the output \ac{SI-SDR} was 12.11[dB] and the \ac{WER} 15.3\% while Conv-Tasnet only achieved 4.2[dB] and
 43.75\%, respectively.

\vspace{-1.3\baselineskip}
\subsubsection{Online Mode}
To facilitate low-latency implementation, we applied the algorithm to 3~sec long segments with a 1~sec look-ahead.
The results for online mode are depicted in the lower part of  Table~\ref{table:online_offline_results}, indicating only a slight degradation compared to batch mode.
\vspace{-1.3\baselineskip}
\subsubsection{Analysis of \ac{VAD} Results}
The \ac{VAD} results were compared to a simple energy-based \ac{VAD} and to WebRTC VAD network. In the energy-based \ac{VAD}, each T-F bin at the output of the separation network (the mask) is compared with an activation threshold and classified as an active or non-active bin. For each time frame, if the number of active bins exceeds a  threshold, the frame is declared a speech-active frame. We examined several threshold pairs and selected the most suitable pair based on the best \ac{VAD} accuracy. We used $T_a=0.3$ and $T_s=0.25$, where $T_a$ is the T-F bin activation threshold and $T_s$ is the threshold for the number of bins to declare speech-active frame. 
WebRTC\footnote{Available online at \url{https://webrtc.org/}}
is a project providing real-time communication capabilities for many different applications. The \ac{VAD} 
 implemented in WebRTC is reportedly one of the best \acp{VAD} available. In setting this \ac{VAD}, we used a latency of 30 ms and multiple frequency band features
with a pre-trained GMM classifier.
Using the new simulated dataset, our embedded \ac{VAD} achieved accuracy, recall, and precision measures of  $94\%,0.96\%,0.94\%$, the energy-based \ac{VAD} achieved $73\%,0.96\%,0.73\%$, and the WebRTC \ac{VAD} achieved $84\%,0.99\%,0.8\%$, respectively. We can, therefore, conclude that incorporating the \ac{VAD} into the separation scheme is beneficial in terms of detection accuracy, although it comes with a slight performance degradation in the speaker separation measures.


\section{Conclusions}
\label{sec:conc}
We presented a speaker separation algorithm that employs a \ac{TCN} module. 
Our results indicate that our model performs comparably or
better than current state-of-the-art separation methods, with the added benefits of success in real-world scenarios. A comprehensive experimental study with both simulated and real recordings
supports our claims.
%
Furthermore, we have demonstrated the effectiveness of using the \ac{STFT}-\ac{iSTFT} instead of the learnable encoder and decoder as in \cite{cord2022monaural}. This approach, which leverages the compact \ac{STFT} representation, proves advantageous in minimizing memory consumption. It should be noted that selecting an adequately large window size is crucial to ensure stable performance in the presence of reverberation.

A new simulated dataset incorporating more realistic adverse acoustic conditions was also introduced. We also comprehensively analyzed the results, considering various parameters such as speakers' gender, signal-to-noise ratio (SNR) level, reverberation level, and distance between the speakers and the microphone.
The method was also applied to real-world data recorded on a humanoid robot in our acoustic laboratory at BIU. By applying our method to this dataset, we demonstrate its superior performance compared to competing methods in a challenging and realistic scenario in both \ac{SI-SDR} and \ac{WER} measures. Our separation algorithm was tested with NVIDIA's ASR system and significantly improved the \ac{WER} in mid-level reverberation, $T_{60}\approx 350$~ms, and low noise levels. The overlap between speakers was set in a realistic range, imitating normal conversations, as fully overlapping utterances are unrealistic.

We note that in the current audio processing architecture, two independent audio streams are simultaneously transcribed by the \ac{ASR} system. For later use by the dialogue system, we may also use our \ac{VAD} decisions to preserve the time consistency of the transcribed speech signals. The \ac{VAD} can also prove beneficial in applying a subsequent processing stage, e.g., multi-microphone beamforming. The embedded \ac{VAD} outperforms both na\"{i}ve and state-of-the-art \acp{VAD}.


\vspace{-11pt}
\section*{Acknowledgments}
We thank Mr. Pini Tandeitnik and Mr. Yoav Elinson for their professional assistance during the acoustic room setup, the recordings, and the ASR application. We also thank Yoav Elinson for thoroughly analyzing the reverberation level in the WHAMR! dataset signals.

\balance
\bibliographystyle{ieeetr}
\bibliography{refs.bib}

\begin{thebibliography}{10}

\bibitem{hershey2016deep}
J.~R. Hershey, Z.~Chen, J.~Le~Roux, and S.~Watanabe, ``Deep clustering: Discriminative embeddings for segmentation and separation,'' in {\em IEEE international conference on acoustics, speech and signal processing (ICASSP)}, pp.~31--35, 2016.

\bibitem{yu2017permutation}
D.~Yu, M.~Kolb{\ae}k, Z.-H. Tan, and J.~Jensen, ``Permutation invariant training of deep models for speaker-independent multi-talker speech separation,'' in {\em IEEE International Conference on Acoustics, Speech and Signal Processing (ICASSP)}, pp.~241--245, 2017.

\bibitem{luo2019conv}
Y.~Luo and N.~Mesgarani, ``Conv-tasnet: Surpassing ideal time--frequency magnitude masking for speech separation,'' {\em IEEE/ACM transactions on audio, speech, and language processing}, vol.~27, no.~8, pp.~1256--1266, 2019.

\bibitem{luo2020dual}
Y.~Luo, Z.~Chen, and T.~Yoshioka, ``Dual-path {RNN}: efficient long sequence modeling for time-domain single-channel speech separation,'' in {\em IEEE International Conference on Acoustics, Speech and Signal Processing (ICASSP)}, pp.~46--50, 2020.

\bibitem{zeghidour2021wavesplit}
N.~Zeghidour and D.~Grangier, ``Wavesplit: End-to-end speech separation by speaker clustering,'' {\em IEEE/ACM Transactions on Audio, Speech, and Language Processing}, vol.~29, pp.~2840--2849, 2021.

\bibitem{zhao2023mossformer}
S.~Zhao and B.~Ma, ``Mossformer: Pushing the performance limit of monaural speech separation using gated single-head transformer with convolution-augmented joint self-attentions,'' in {\em IEEE International Conference on Acoustics, Speech and Signal Processing (ICASSP)}, 2023.

\bibitem{nachmani2020voice}
E.~Nachmani, Y.~Adi, and L.~Wolf, ``Voice separation with an unknown number of multiple speakers,'' in {\em International Conference on Machine Learning (ICML)}, pp.~7164--7175, 2020.

\bibitem{lutati2022sepit}
S.~Lutati, E.~Nachmani, and L.~Wolf, ``{SepIt} approaching a single channel speech separation bound,'' {\em arXiv preprint arXiv:2205.11801}, 2022.

\bibitem{le2019sdr}
J.~Le~Roux, S.~Wisdom, H.~Erdogan, and J.~R. Hershey, ``{SDR}--half-baked or well done?,'' in {\em IEEE International Conference on Acoustics, Speech and Signal Processing (ICASSP)}, pp.~626--630, 2019.

\bibitem{subakan2021attention}
C.~Subakan, M.~Ravanelli, S.~Cornell, M.~Bronzi, and J.~Zhong, ``Attention is all you need in speech separation,'' in {\em IEEE International Conference on Acoustics, Speech and Signal Processing (ICASSP)}, pp.~21--25, 2021.

\bibitem{subakan2022using}
C.~Subakan, M.~Ravanelli, S.~Cornell, F.~Grondin, and M.~Bronzi, ``On using transformers for speech-separation,'' {\em arXiv preprint arXiv:2202.02884}, 2022.

\bibitem{tzinis2022compute}
E.~Tzinis, Z.~Wang, X.~Jiang, and P.~Smaragdis, ``Compute and memory efficient universal sound source separation,'' {\em Journal of Signal Processing Systems}, vol.~94, no.~2, pp.~245--259, 2022.

\bibitem{wichern2019wham}
G.~Wichern, J.~Antognini, M.~Flynn, L.~R. Zhu, E.~McQuinn, D.~Crow, E.~Manilow, and J.~L. Roux, ``{WHAM!: Extending Speech Separation to Noisy Environments},'' in {\em Proc. Interspeech}, pp.~1368--1372, 2019.

\bibitem{cord2022monaural}
T.~Cord-Landwehr, C.~Boeddeker, T.~Von~Neumann, C.~Zoril{\u{a}}, R.~Doddipatla, and R.~Haeb-Umbach, ``Monaural source separation: From anechoic to reverberant environments,'' in {\em International Workshop on Acoustic Signal Enhancement (IWAENC)}, 2022.

\bibitem{heitkaemper2020demystifying}
J.~Heitkaemper, D.~Jakobeit, C.~Boeddeker, L.~Drude, and R.~Haeb-Umbach, ``Demystifying {TasNet}: A dissecting approach,'' in {\em IEEE International Conference on Acoustics, Speech and Signal Processing (ICASSP)}, pp.~6359--6363, 2020.

\bibitem{wang2021continuous}
D.~Wang, T.~Yoshioka, Z.~Chen, X.~Wang, T.~Zhou, and Z.~Meng, ``Continuous speech separation with ad hoc microphone arrays,'' in {\em 2021 29th European Signal Processing Conference (EUSIPCO)}, pp.~1100--1104, IEEE, 2021.

\bibitem{lin2021sparsely}
Q.~Lin, L.~Yang, X.~Wang, L.~Xie, C.~Jia, and J.~Wang, ``Sparsely overlapped speech training in the time domain: Joint learning of target speech separation and personal vad benefits,'' in {\em 2021 Asia-Pacific Signal and Information Processing Association Annual Summit and Conference (APSIPA ASC)}, pp.~689--693, IEEE, 2021.

\bibitem{maciejewski2020whamr}
M.~Maciejewski, G.~Wichern, E.~McQuinn, and J.~Le~Roux, ``{WHAMR!}: Noisy and reverberant single-channel speech separation,'' in {\em IEEE International Conference on Acoustics, Speech and Signal Processing (ICASSP)}, pp.~696--700, 2020.

\bibitem{lea2016temporal}
C.~Lea, R.~Vidal, A.~Reiter, and G.~D. Hager, ``Temporal convolutional networks: {A} unified approach to action segmentation,'' in {\em European Conference on Computer Vision (ECCV)}, (Amsterdam, The Netherlands), pp.~47--54, Springer, Oct. 2016.

\bibitem{wang2022stft}
Z.-Q. Wang, G.~Wichern, S.~Watanabe, and J.~Le~Roux, ``{STFT}-domain neural speech enhancement with very low algorithmic latency,'' {\em IEEE/ACM Transactions on Audio, Speech, and Language Processing}, vol.~31, pp.~397--410, 2022.

\bibitem{ravenscroft2023deformable}
W.~Ravenscroft, S.~Goetze, and T.~Hain, ``Deformable temporal convolutional networks for monaural noisy reverberant speech separation,'' in {\em IEEE International Conference on Acoustics, Speech and Signal Processing (ICASSP)}, 2023.

\bibitem{9966661}
Q.~Zhang, X.~Qian, Z.~Ni, A.~Nicolson, E.~Ambikairajah, and H.~Li, ``A time-frequency attention module for neural speech enhancement,'' {\em IEEE/ACM Transactions on Audio, Speech, and Language Processing}, vol.~31, pp.~462--475, 2023.

\bibitem{liu2020rethinking}
F.~Liu, X.~Ren, Z.~Zhang, X.~Sun, and Y.~Zou, ``Rethinking skip connection with layer normalization,'' in {\em Proceedings of the 28th international conference on computational linguistics}, pp.~3586--3598, 2020.

\bibitem{kolbaek2017multitalker}
M.~Kolb{\ae}k, D.~Yu, Z.-H. Tan, and J.~Jensen, ``Multitalker speech separation with utterance-level permutation invariant training of deep recurrent neural networks,'' {\em IEEE/ACM Transactions on Audio, Speech, and Language Processing}, vol.~25, no.~10, pp.~1901--1913, 2017.

\bibitem{yemini2020scene}
Y.~Yemini, E.~Fetaya, H.~Maron, and S.~Gannot, ``Scene-agnostic multi-microphone speech dereverberation,'' in {\em Proc. of Interspeech}, (Brno, The Czech Republic), 2021.

\bibitem{delcroix2014linear}
M.~Delcroix, T.~Yoshioka, A.~Ogawa, Y.~Kubo, M.~Fujimoto, N.~Ito, K.~Kinoshita, M.~Espi, T.~Hori, T.~Nakatani, {\em et~al.}, ``Linear prediction-based dereverberation with advanced speech enhancement and recognition technologies for the reverb challenge,'' in {\em Reverb workshop}, 2014.

\bibitem{panayotov2015librispeech}
V.~Panayotov, G.~Chen, D.~Povey, and S.~Khudanpur, ``Librispeech: an asr corpus based on public domain audio books,'' in {\em IEEE International Conference on Acoustics, Speech and Signal Processing (ICASSP)}, pp.~5206--5210, 2015.

\bibitem{allen1979image}
J.~B. Allen and D.~A. Berkley, ``Image method for efficiently simulating small-room acoustics,'' {\em The Journal of the Acoustical Society of America}, vol.~65, no.~4, pp.~943--950, 1979.

\bibitem{habets2006room}
E.~A. Habets, ``Room impulse response generator,'' tech. rep., Friedrich-Alexander-Universit\"{a}t Erlangen-N\"{u}rnberg, 2014.

\bibitem{tzinis2020sudo}
E.~Tzinis, Z.~Wang, and P.~Smaragdis, ``Sudo rm-rf: Efficient networks for universal audio source separation,'' in {\em IEEE International Workshop on Machine Learning for Signal Processing (MLSP)}, 2020.

\end{thebibliography}

\end{document}